**Direct observation of a highly spin-polarized organic spinterface at room temperature**

F. Djeghloul[1], F. Ibrahim[1], M. Cantoni[2], M. Bowen[1*], L. Joly[1,3], S. Boukari[1], P. Ohresser[4], F. Bertran[4], P. Lefèvre[4], P. Thakur[5], F. Scheurer[1], T. Miyamachi[6], R. Mattana[7], P. Seneor[7], A. Jaafar[1], C. Rinaldi[2], S. Javaid[1], J. Arabski[1], J.-P. Kappler[1], W. Wulfhekel[6], N.B. Brookes[5], R. Bertacco[2], A. Taleb-Ibrahimi[4], M. Alouani[1], E. Beaurepaire[1], W. Weber[1]

1. IPCMS, Université de Strasbourg, 23 rue du Loess BP 43 67034 Strasbourg
2. LNESS – Dipartimento di Fisica, Politecnico di Milano, Via Anzani 42, 22100 Como, Italy;
3. Swiss Light Source, Paul Scherrer Institut, 5232 Villigen PSI, Switzerland
4. Synchrotron Soleil, L'Orme des Merisiers Saint-Aubin - BP 48 91192 Gif-sur-Yvette Cedex
5. European Synchrotron Radiation Facility (ESRF), 38043 Grenoble Cedex, France
6. Physikalisches Institut and Center for Functional Nanostructures, Karlsruhe Institute of Technology, Wolfgang-Gaede-Strasse 1, 76131 Karlsruhe, Germany
7. Unité Mixte de Physique CNRS/Thales, 91767 Palaiseau France associée à l'Université de Paris-Sud, 91405 Orsay, France ;

**The design of large-scale electronic circuits that are entirely spintronics-driven requires a current source that is highly spin-polarised at and beyond room temperature, cheap to build, efficient at the nanoscale and straightforward to integrate with semiconductors. Yet despite research within several subfields spanning nearly two decades, this key building block is still lacking. We experimentally and theoretically show how the interface between Co and phthalocyanine molecules constitutes a promising candidate. Spin-polarised direct and inverse photoemission experiments reveal a high degree of spin polarisation at room temperature at this interface. We measured a magnetic moment on the molecule's nitrogen π orbitals, which substantiates an ab-initio theoretical description of highly spin-polarised charge conduction across the interface due to differing spinterface formation mechanisms in each spin channel. We propose, through this example, a recipe to engineer simple organic-inorganic interfaces with remarkable spintronic properties that can endure well above room temperature.**

Technological progress in the past decade has been nothing short of astounding as revealed by our maturing information society. An important milestone will be to design not only electrical components but entire circuits that pervasively utilize the electron spin as well as its charge. In this vein, research has focused on the interface between ferromagnets (FM), whose current is spin-polarised, and organic semiconductors (OS), which have been identified as a promising medium to transport spin-encoded information due to low spin-orbit induced spin decoherence in this class of semiconductors[1]. When integrated into devices, such interfaces can yield large values of magnetoresistance at low temperature, whether in the diffusive regime[2], in the ballistic regime across individual molecules[3] or in the tunneling regime[4]. As supported by a phenomenological model, this latter result could underscore how, due to molecular chemisorption onto a transition metal surface, the OS's molecules at the interface may exhibit a molecular orbital (MO) at the Fermi level $E_F$[5] that extends the electrode conduction onto the first molecular layer[6]. Due to exchange-split bands, the unequal density of states (DOS) of the two spin populations at $E_F$ in the FM is then believed to lead to a spin-selective broadening of this MO[4], i.e. to a spin-polarised interface[6] that is termed a spinterface[7]. This original mechanism of spinterface formation leads to band-induced spinterface states (BISS). However, experiments have thus far not revealed large values of room temperature (RT) spin polarization (P) at such FM/OS interfaces, whether through spectroscopy techniques[8,9] or on actual devices[10]. In this sense, a validation of the promise behind the spinterface concept[4,7] --- namely more efficient interfaces for spintronic applications --- is still lacking.

---
* email: bowen@unistra.fr

In what follows, we demonstrate that solving this riddle requires the study of FM/OS interfaces whose structure and electronic properties are well characterised. Given the link between photoemission (PE) and magnetotransport spectroscopy techniques[11], we have performed spin-polarised direct and inverse PE experiments at RT on interfaces between fcc Co(001) and MnPc (see the molecular schematic in the inset to fig. 2e) or $H_2Pc$ as potential spinterface candidates[6,3]. The PE experiments reveal the presence of Pc-induced states close to $E_F$[8,9]. In order to extract the signal coming only from the molecular sites, we adopt a subtraction procedure that takes into account the attenuation of the signal arising from ever deeper atomic sites away from the sample surface (see SI). We present in Fig. 1a the spin-resolved difference spectra of direct and inverse PE spectroscopy of Co/MnPc at RT (2.6 ML MnPc for direct and 2 ML MnPc for inverse PE) that are obtained by this subtraction procedure. Both direct and inverse PE experiments reveal significant (nearly no) spin ↑ (↓) intensity at/near $E_F$, which indicates a high P of the Pc-induced states in the vicinity of $E_F$. We note that very similar difference spectra (in direct PE) are also obtained in the case of $H_2Pc$, which shows that the central $Mn^{2+}$-ion in MnPc plays a minor role in the formation of the spinterface. Assuming that the spin asymmetry of spectra is directly related to P (see eg 11), we can safely state that the RT P at $E_F$ of the first two layers of MnPc or $H_2Pc$ adsorbed on Co(001) reaches at least +80%, ie is opposite in sign to that of bare Co.

We now confirm the interfacial nature of P by examining the impact of additional Pc coverage. Upon appropriately subtracting the spin-resolved spectra of 1 ML $H_2Pc$/Co from those of 2 ML $H_2Pc$/Co, the intensity of the interface states is strongly reduced (see Fig. 1b): the second Pc layer contributes only 20 % to the total intensity of the interface states of Fig. 1(a), which could reflect deviations from perfect layer-by-layer growth. The third ML does not contribute at all to the interface states' intensity. We have also excluded the artefact of an altered Co interface magnetism on our analysis and conclusions (see SI).

To determine whether these interface states originate dominantly from the Co substrate or the Pc-overlayer, we compared data for photon energies of 20 eV and 100 eV (see Fig. 1c). From 20eV to 100eV, the cross section of photoionisation for free atoms decreases by over one order of magnitude for 2$p$ states (C and N) while that for 3$d$ states (Co and Mn) does not vary much[12]. We expect that such a large effect for free atoms shall trend similarly in solid-state systems. Consequently, if the interface states were mainly of Co 3d character, they should also be present at 100 eV photon energy. However, the spin-resolved direct PE difference spectra at 100 eV show no indication for any Pc-induced structure at low binding energies. We thus conclude that the interface states are mainly of C or N 2$p$ character.

Why does the interface between fcc Co(001) and MnPc or $H_2Pc$ exhibit such a high P of PE at $E_F$, and this at RT? We propose the following key extension to the spinterface concept[4,7]: *highly efficient, thermally robust spinterfaces may be engineered by choosing the ferromagnet/molecule pair such that A) the dominant interfacial hybridization mechanism involves states at/near $E_F$ from the ferromagnet (FM) and molecule that are present only in*

*one spin channel so as to B) promote, in addition to the well-known formation of BISS states[3,4,7] in that spin channel, the hybridization in the other spin channel between the FM's surface states at the vicinity of $E_F$ and MOs of the molecule. This mechanism ensures that energetically narrow and strongly spin-polarized hybrid interface states, called surface-induced spinterface states (SISS), are pinned close to the Fermi level so as to drive the interface's spintronic response. The resilience of the ensuing spinterface properties against thermal disorder are enhanced not only by a large FM Curie temperature but also when C) direct exchange coupling that results from the hybridization mechanism magnetises at least some of the molecule's atoms.*

When considering all electronic orbitals, ab-initio calculations on Co/Pc interfaces with unrelaxed atomic positions reveal a P that can reach 25% per atomic site of the molecule[6]. To more realistically describe the interface, our formalism now relaxes atomic positions and includes van der Waals forces so as to quantitatively reproduce the crucially important molecule-substrate distance inferred from x-ray standing wave measurements. This leads to a final distance $\Delta z$ between Co and the adsorbed molecule of 2.1Å.

To unravel the formation of the spinterface, we first consider the 'molecule-Co' system as calculated using the actual atomic positions of the final interface, but we artificially impose $\Delta z$=6.6Å. We can then examine the states of the two systems with a common chemical potential in the absence of interactions between them (see Fig. 2a). The Co d- spin ↓ band crosses $E_F$, while the d spin ↑ band ends at $E-E_F$=-0.7eV. Above this energy level, the spin ↑ sub-band exhibits only small DOS spikes that correspond to surface states. We note in particular one surface state at $E_F$ with a strong perpendicular component (z-DOS, black) compared to its planar counterpart (pl-DOS, magenta). We emphasize that these surface states also exhibit a *s*-component of DOS (gray). Near $E_F$, the molecule exhibits a MO only in the spin spin ↓ channel. Adsorption-induced displacements of the molecule's atoms overall promote a slight energy shift (~30meV) of the MOs.

We now turn on interactions between the molecule and the Co surface by reducing $\Delta z$ to 3.5Å (fig. 2b). At this distance, π orbitals that spatially extend perpendicularly to the nascent interface promote wavefunction overlap between the molecule and Co surface sites, causing $E_F$ to shift from E=-2.4eV to E=-2.2eV. At the vicinity of $E_F$, the Co spin ↓ states and spin ↑ surface states are little affected. In contrast, the interaction strongly modifies the molecule's states: while planar states remain mostly unaffected, perpendicular states experience the onset of hybridization. In particular, this results in the energy dispersion of the initially sharp spin ↓ states in Fig. 2a at -2.4eV and -2.2eV. We emphasize here that there are no spin ↑ MO at/near $E_F$ at $\Delta z$=3.5Å (right-hand panel of fig. 2b).

At the final $\Delta z$=2.1Å (fig. 2c), the molecule and Co surface sites may fully hybridize to form the spinterface. More precisely, all combinations of *s-p*, *p-d* and *s-d* hybridization may occur. Although fcc Co(001) has, near $E_F$, no *p* states and a highly spin-polarized *d* band, the flat, spin-degenerate *s*-band that crosses $E_F$ is essentially responsible, through *s-d* hybridization[13], for the only moderate 45% spin polarization of conduction electrons. Yet, referring to Fig. 2c, the spinterface formation involves Co *s*-states (gray datasets) only very weakly. Thus, although fcc Co(001) is obviously not half-metallic[14,15], the Co/MnPc spinterface shall strongly transmit the highly spin-polarized *d*-component of the Co DOS and attenuate the *s* and *p* components.

How is the Co $d$-band DOS transmitted onto the molecule in each spin channel? Prior to adsorption and in the spin ↓ channel, the Co $d$ band z-DOS intersects $E_F$ and the z-DOS of the free molecule also exhibits a MO at/near $E_F$. Hybridization is therefore governed by the well-known spinterface mechanism of spin-dependent broadening of MOs due to band hybridisation[4,7,3]. The resulting BISS (band-induced spinterface states) are shaded in red in fig. 2d. These BISS exhibit a flat, continuous energy dependence across $E_F$.

However, the molecule does not exhibit any sizeable, preexisting spin ↑ z-DOS at the vicinity of $E_F$ to hybridise with, and the Co surface's $d$-band doesn't cross $E_F$. Another spinterface formation mechanism must therefore account for the appearance of entirely new, hybrid states in the spin ↑ channel within -2.7eV< E < -1.9eV, i.e. at the vicinity of $E_F$, (see right-hand panel of Fig. 2c and the segment of the spinterface z-DOS shaded in green in fig. 2d). We propose that preexisting Co surface states (see left-hand panel of Fig.2a and b) pin initially distant MOs to $E_F$. The narrow energy width of these surface-induced spinterface states (SISS) reflects that of both the preexisting Co surface states --- because the surface atoms are missing bonds --- and of the preexisting MOs. Due to the Pauli exclusion principle, these newly formed SISS cannot occupy the spin ↓ states since they are already occupied by Co, and hence appear only in the spin ↑ channel. The presence of two sharp, tall peaks near $E_F$ reflects a lifting of degeneracy induced by upward(downward) buckling of the benzene rings below(at) $E_F$ along each of the two orthogonal axes that define the free molecule's 4-fold symmetry. This underscores how crucial it is to fully relax the interface structure if one wishes to study SISS.

Since surface states naturally lie at the vicinity of $E_F$, so shall SISS. Although SISS may appear as energetically sharp DOS peaks, which could reflect localization, SISS contribute to conduction across the interface. Indeed, the spectral signature of the SISS appears in the spin ↑ z-DOS of both Co surface and molecular sites (compare graphs of fig. 2c). Focusing now on the DOS that contributes to transport at RT, we present in Fig. 3c-d spin-polarised spatial maps, taken along the dashed line of Fig. 3a, of the Co/MnPc interface DOS within $E_F$ -25 meV < E < $E_F$+25 meV (see Fig. 3b). Aside from the central Mn site, the remaining N and C sites exhibit very large positive P at $E_F$ thanks to electronic states that are clearly hybridised with the Co interface atoms. In fact, these interface states are present on all atomic species of the molecule (fig. 2e) and their amplitude trends with the number of molecular nearest-neighbours for a given Co spinterface site.

At $E_F$, both the energetically smooth BISS in the spin ↓ channel and the energetically sharp SISS in the spin ↑ channel define the sign and amplitude of the spinterface's P. Due in large part to the energetically sharp SISS that crosses $E_F$, we find that P=80%. Thus, considering the limitations of the comparison, we find that both theory and the direct/inverse PE experiments yield the same sign and high amplitude of P at $E_F$ (see fig. 1a and 2e). Furthermore, peaks in the spin ↑ (↓) PE (see fig. 1a) and DOS spectra (see fig. 2d) at ~E- $E_F$ =-0.3(-1.0) eV underscore a reasonably good agreement between theory and the direct PE experiment thanks to its good energy resolution (130meV), while a qualitative agreement is found with inverse PE.

Since both PE experiments and ab-initio theory describe how the molecule's sites are spin-polarised, we now consider the magnetic properties of the spinterface. Referring to the on-site local magnetisation density map of Fig. 4a, our theory indicates that a strong antiferromagnetic (AF) coupling between Co and the numerous C benzene sites leads to a total magnetic moment for all C atoms of -0.22 $\mu_B$. Within a Hund's rule description, this is expected since the Co *d* orbitals are more than half-filled. Only the partially filled $d\downarrow$ band may then hybridize, so that the coupling between C and Co is mediated essentially by minority electrons. Direct *p-d* coupling then leads to an exchange splitting of the C majority and minority DOS that is opposite in direction to that of Co.

The magnetic coupling of N sites is more subtle. Indeed, although N is coupled AF to Mn for free MnPc, molecular adsorption onto Co leads, through *d-d* hybridization, to F coupling between Mn and Co (as expected since the Mn *d* band is less than half filled)[16,6]. Due to aromaticity, this F coupling is found to drive F coupling onto all N and C pyrrole sites. Thus, although C and N sites both contribute to the high P at $E_F$, their magnetisations are in fact opposite to one another.

If the molecule z-DOS is spin-polarized at EF owing to BISS and SISS, then the molecule's π DOS at EF should be spin-polarised. To support this theoretical description of spinterface magnetism, and as a tenet of spintronically active interfaces[31], we have performed x-ray magnetic circular dichroism (XMCD) experiments at the N *K* edge of MnPc's 8 nitrogen sites (see Methods). Referring to Fig. 4b, we witness XMCD intensity within the energy range corresponding to final 2p π (i.e. that probe the z-DOS just above $E_F$), but not 2p σ, states[6]. This unequivocal XMCD signal is very strong compared to the stray XMCD signal obtained when MnPc is adsorbed onto Cu(001) (see Fig. 4c), for which one does not expect the presence of on-site magnetic moments. The sharp absorption peak at 401eV in the Cu/MnPc spectrum, which leads to the derivative-like XMCD signal, is in fact due to low-temperature $N_2$ adsorption. Since these are *K* edge transitions, we can only state[18] that an orbital magnetic moment appears on the final N 2p π states at the Co/MnPc spinterface, the sign of which is in agreement with that found theoretically. This experimentally confirms that the N z-DOS is spin-polarised as we have described theoretically.

We now discuss spintronics prospects for these Co/Pc spinterfaces. Indeed, an ideal spin-polarized current source (IspCS) should 1) exhibit a very high degree of spin polarisation P that 2) endures well above RT for technological applications; 3) be both cheap and straightforward to synthesize considering existing industrial capabilities; 4) be compatible with miniaturisation challenges at the nanoscale; and 5) provide an easy integration path with a semiconductor so as to enable transport of, and operations on, the highly spin-polarised current. Behind criterion 5 lies the original promise of the spintronics field to promote the rise of an electronics in which not only individual electronic components (eg read heads in hard disks) but entire electronic circuits are conceived so as to encode and transport information using the electron spin.

Candidates toward an IspCS include half-metallic ferromagnets, which ideally conduct electrons of only one spin direction[14] and could, using merely a band-hybridisation mechanism of spinterface formation[4,7], lead to efficient spinterfaces. Such materials have been studied using direct PE[19] and been integrated into devices with sizeable P, not only at low temperature[15] but also at RT[20]. However, this track fails criteria 3 and 4 for an IspCS because such materials are sensitive to disorder. Dilute magnetic semiconductors offer an interesting solution to criterion 5, but lose their half-metallic property well below RT[21]. Another track is to resistively filter the current so as to spin-polarise it. Fe/MgO-based IspCS accomplish this[22] through tunneling across MgO[23] and can reach P = 85 %[24], but this resistive solution to spin-filtering a) must involve several dielectric monolayers that b) must be of finite lateral extent in order to promote $k_{//}$ conservation. This resistive solution is therefore not as practical toward nanoscale applications (criterion 4) as a conductive one involving merely an interface that can scale down laterally to the individual molecule[3].

In contrast, the Co/Pc interface involves differing spinterface formation mechanism in each spin channel to yield a high P (criterion 1). Since both mechanisms are driven by direct rather than indirect[16] hybridisation, the resulting current source is spin-filtered across a conductive[5,6] interface (criterion 4) and inherits the large temperature resiliency of the Co interface magnetisation (criterion 2). Such spinterfaces utilize cheap, abundant materials that can be straightforwardly deposited and will not degrade when processed appropriately into devices[25] even at typically large process temperatures[26] (criterion 3). Finally, with its spin-polarized molecular plane, this IspCS candidate elegantly mitigates the conductivity mismatch problem[27] associated with interfaces between metals and semiconductors, which is promising toward satisfying criterion 5. Indeed, the hybridization of wavefunctions from the interfacial molecular plane of high P with those of subsequent molecular layers away from the interface is intrinsically favored. Furthermore, referring to Fig. 3, conductivity is substantially lowered when going from Co to the Pc spinterface due to a strongly attenuated spin ↓ channel.

Thus, our work on Co/Pc interfaces provides a direct proof of the promise behind the spinterface concept, which was initially described in terms of band-induced spinterface states (BISS)[4,7]. We propose to extend this concept to include the additional spinterface formation mechanism of surface-induced spinterface states (SISS). SISS appear if the FM band of the dominant hybridisation mechanism is absent near $E_F$ in one spin channel. This criterion is for example satisfied in the spin ↑ channel by strong ferromagnets such as Co or Ni. By combining BISS and SISS in separate spin channels, the spintronic response of these spinterfaces is not only large but can potentially be controlled through external stimuli. For example, we find that rotating the magnetisation by 90° shifts the SISS peak at $E_F$ by ~1meV, leading to a 10% change in P. Underscoring this effect is the spinterface's magnetic anisotropy, which can itself in principle be controlled using an electric field (eg 28) so as to more substantially alter the spinterface properties. Finally, these spinterfaces constitute a strong candidate toward satisfying the five criteria for an IspCS[29], so as to pervasively use the electron spin, not simply in individual electronic components, but in future electronics industrial designs. Beyond future Co/Pc-based spintronic demonstrators based on the well-established tunneling mechanism of spin-polarized transport, we are presently working to extend these spinterface-induced IspCS concepts to memristive organic interfaces[30], so as to pave the way for robust organic multifunctional devices alongside their inorganic counterparts[31].

**Methods**

To prepare samples for x-ray absorption (XAS), spin-polarised photoemission (SPARPES) and spin-polarised inverse photoemission (SPIPES) experiments, we used a Cu(100) single crystal as substrate. It was cleaned by sputtering and annealing at 900 K. MnPc and $H_2$Pc were sublimated (P~$10^{-9}$ mbar, 1 monolayer (ML)=0.38 nm) so as to form ultrathin films on Cu(100) or on Co(100) layers epitaxially grown on Cu(001). XAS were acquired (beamlines SIM at SLS and ID8 at ESRF) in total electron yield mode (P<$2 \times 10^{-10}$mbar) by reversing both the circular polarity of the photons and the sign of the external magnetic field. XAS were measured at the N K edge. The XMCD signal (ID8) was normalized to the height of the absorption edge step. The incidence angle was ~50° to be sensitive to both in- and out-of-plane orbitals. We affirm a successful subtraction of the Co $L_{3,2}$ harmonics from the N K edge XMCD. Indeed, the N K edge XMCD is of same sign as the remnant Co $L_3$ harmonic. Since the Co $L_3$ and $L_2$ harmonics are necessarily of opposite sign, the measured XMCD cannot arise from the Co $L_2$ harmonic. Note that beamline ID8 exhibits a strong C absorption within the background spectrum that precluded XAS/XMCD experiments at the C K edge.

SPARPES experiments were undertaken on the Cassiopee Beamline at Synchrotron Soleil using photons at 20 and 100 eV and with the horizontal electric field impinging upon the sample at 45°. Photoelectrons were then acquired along a direction normal to the sample surface. The energy resolution is 130 meV.

SPIPES experiments were performed using a collimated and transversely polarised electron beam with 25% polarisation, from a GaAs photocathode. The SPIPES spectra are taken in the isochromatic mode by collecting photons at a fixed photon energy of 9.3 eV, while varying the incident-beam energy[32]. The energy of the incident electrons was varied between 9 and 17 eV. Data were collected at room temperature and at normal incidence. The energy resolution is 750 meV.

All density functional theory (DFT) calculations were carried by means of the VASP package[33] and the generalized gradient approximation for exchange-correlation potential as parametrized by Perdew, Burke, and Ernzerhof[34]. We used the projector augmented wave (PAW) pseudopotentials as provided by VASP[35]. The van der Waals (vdW) weak interactions were computed within the so called GGA-D2 approach developed by Grimme[36] and later implemented in the VASP package[37]. Our formalism can correctly reproduce the experimentally determined atomic distances between molecular sites and metallic sites. Fcc Co(001) and fcc Cu(001) surfaces were modeled by using a supercell of 3 atomic monolayers of 8x8 atoms separated by a vacuum region. The lattice vector perpendicular to the surface is 3 nm. This results in a supercell of 249 atoms, including the 57 atoms of the MnPc molecule. Since experiments used cobalt epitaxially grown on Cu, we used the fcc lattice parameter of 0.36 nm for both cobalt and copper. We have found that additional monolayers will not change significantly the results[38]. A kinetic energy cutoff of 450 eV has been used for the plane-wave basis set. For our study of a single molecule on metallic surfaces, we used only the gamma point to sample the first Brillouin zone. DOS were calculated using a 1meV energy mesh and a Gaussian broadening of 20meV full-width at half-maximum. Spin-orbit coupling was included pertubatively in the augmentation region at each atomic site.

**Acknowledgements**
F. Djeghloul thanks « Le Ministère de l'enseignement Supérieur et de la Recherche Scientifique d'Algérie » (MESRS) for financial support. We acknowledge financial support from ANR PNANO grants ANR-06-NANO-053-01 and ANR-06-NANO-053-02, the EC Sixth Framework Program (NMP3-CT-2006-033370), the CNRS-PICS Program No. 5275, the Deutsche Forschungsgemeinschaft (DFG), the Center for Functional nanostructures (CFN), the the International Center for Frontier Research in Chemistry and the French German University. S. Javaid thanks the Pakistani government (HEC) for financial support. This work was performed using HPC resources from GENCI-CINES Grant 2012-gem1100.

**Author Contributions**
J.A. purified the molecules. W.We., R. B. and M.B. conceived the photoemission experiments. F.D., M.C., L.J., M.B., S.B., F.B., P.L., C.R., R.B., W.We. and A. T.-I. carried out the photoemission experiments. W.We., L.J., M.C, R.B., S.B., M.B. and E.B. analysed the data. M.B., S.B., E.B., F.S. and W.Wu. conceived the x-ray absorption experiments. S.B., L.J., S.B., P.O., P.T., F.S., M.B., R.M., T.M., S.J., J.-P. K. and E.B carried out the experiments. S.B., P.O., F.S., M.B. and N.B. analysed the data. M.A conceived the ab-initio theory along with M.B., S.B., E.B. and W.We. F.I., S.J. and A.J. carried out the calculations. F.I., M.B., M.A., S.B., W.We. and E.B. analysed the data. M.B. wrote the paper, assisted by F.I., W.We., M.A. S.B., E.B., W.Wu and P.S. M.B. prepared final figures, assisted by W.We., M.C., F.I., P.O. and S.B. All authors discussed the results and commented on the manuscript.

The authors declare no conflict of interest associated with this work. A patent disclosure related to the high-efficiency spinterface invention was recently filed with the CNRS and Université de Strasbourg, disclosure no. DI-04961-01, which has been filed as a provisional patent application.

# Figures

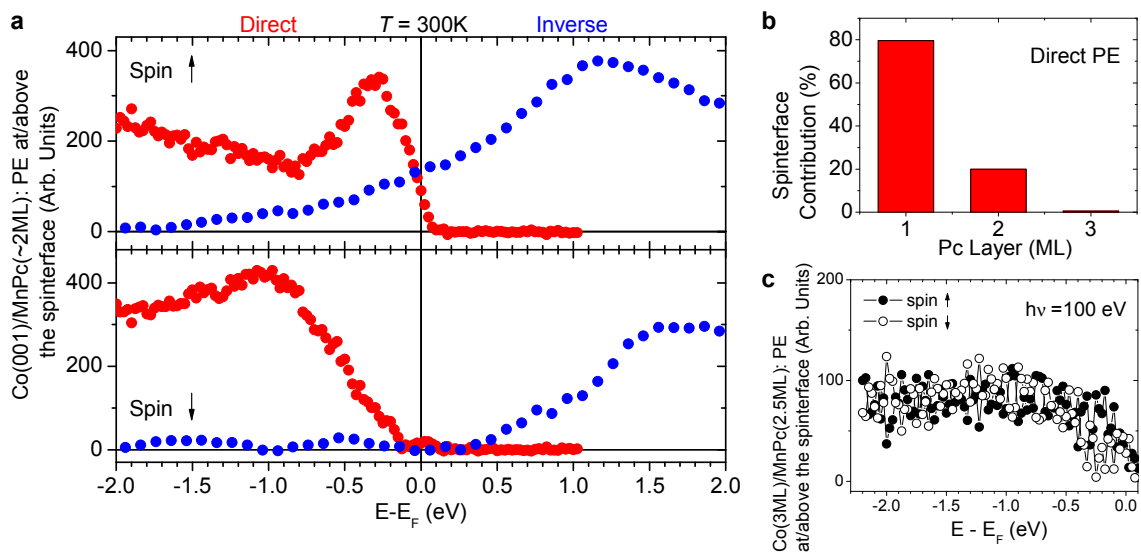

Fig. 1: **Direct and inverse photoemission reveals a high interface spin polarisation P using commonplace Co and phthalocyanine molecules.** (a) Spin-resolved difference spectra of direct (red; hν=20 eV) and inverse (blue) photoemission (PE) spectroscopy at room temperature of Co/MnPc(2.6(2.0) ML for direct(inverse) PE) reveal a P~+80% at $E_F$. (b) The Pc thickness dependence of the direct PE signal (hν=20 eV) reveals that Pc-induced intensity at low binding energies is essentially confined to the interface. (c) Spin-resolved difference spectra of direct PE spectroscopy at room temperature of Co(3 ML)/MnPc(2.6 ML) for hν=100 eV show no sign of any Pc-induced interface state, indicating that the interface states are mainly of C or N 2$p$ character.

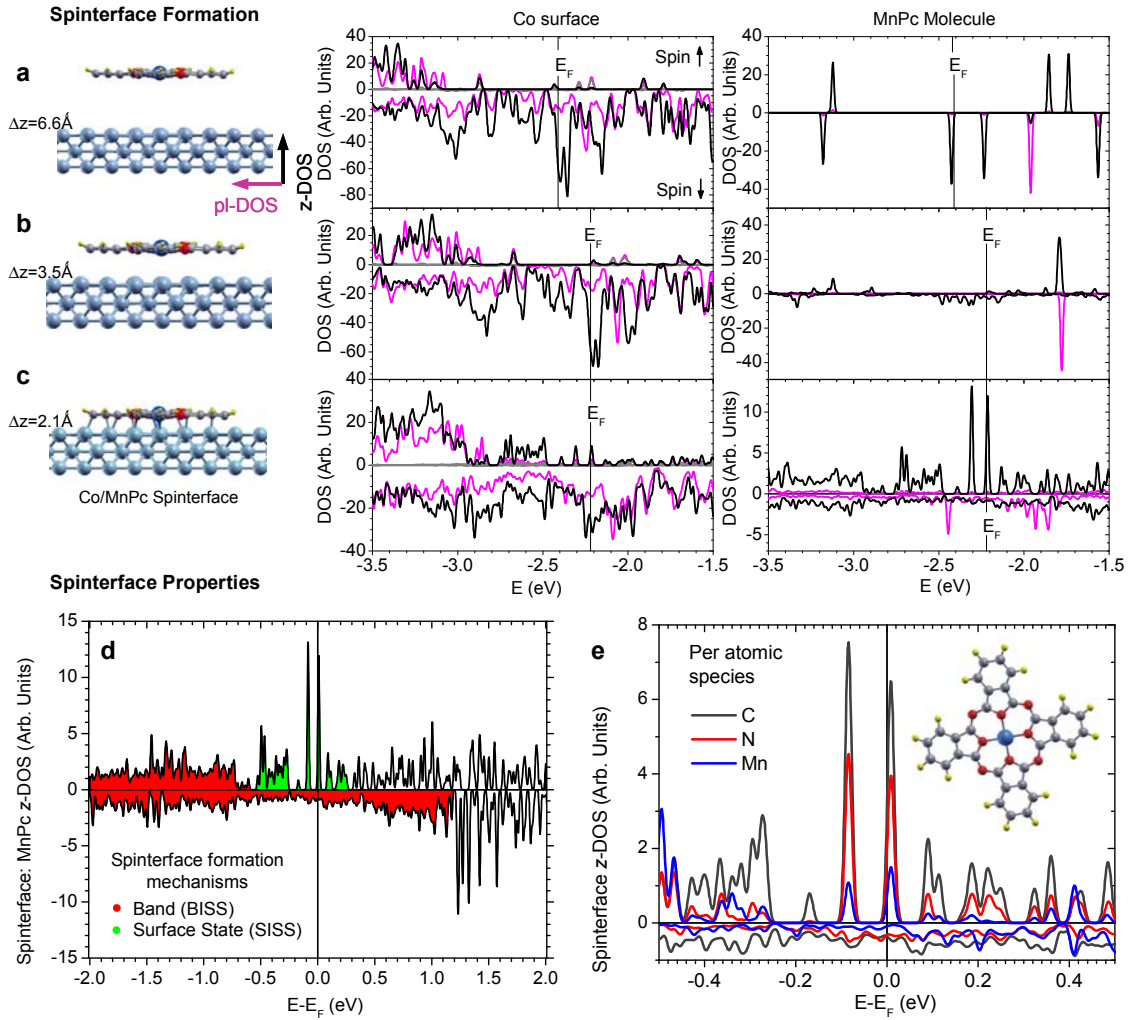

Fig 2: **The formation and properties of the Co/MnPc spinterface reflect distinct mechanisms in each spin channel.** As the distance between molecule and the Co surface is reduced from (a) 6.6Å to (b) 3.5Å and to (c) the final position of 2.1Å, *p-d* hybridization with the Co spin ↓ band causes energetically sharp, spin ↓ MOs in the z-DOS to disperse (red area of panel d), leading to a monotonous spin-↓ z-DOS (black) at/near $E_F$. In the spin ↑ channel at the vicinity of $E_F$, there are neither Co *d* band states nor MOs but simply Co surface states (panel a) that begin to hybridize as the molecule is brought closer in (panel b) and lead, at the final molecular position (panel c), to energetically sharp peaks that cross $E_F$. These surface-induced spinterface states (SISS) carry virtually no Co s-character (gray datasets in panels a,b,c) and involve all atomic species of the molecule (panel e). The spinterface's planar DOS (pl-DOS; magenta) near $E_F$ is mostly featureless and adopts the spin polarization of Co (right- and left-hand graphs of panel c).

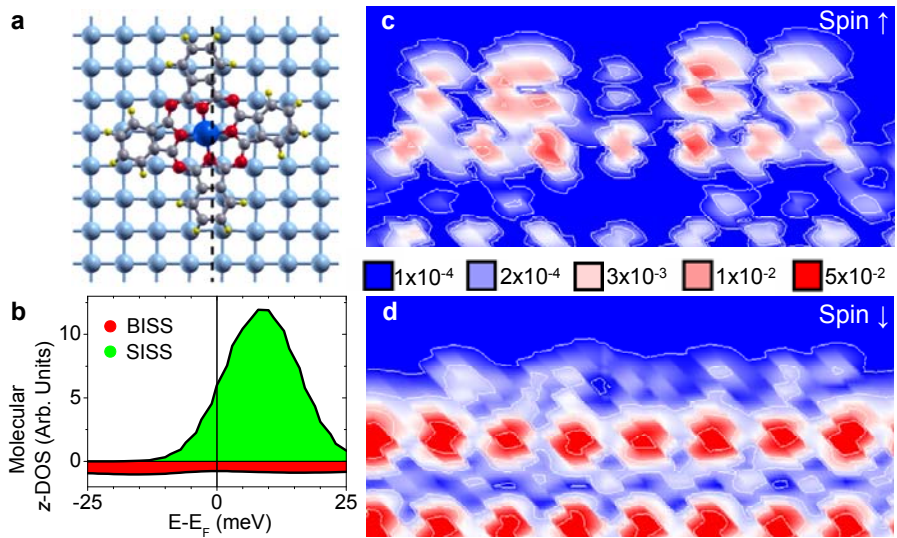

Fig. 3: **The Co/MnPc spinterface as a highly spin-polarised current source.** (a) Adsorption geometry of MnPc on Co(001). The spin ↑ and ↓ z-DOS within $E_F-25 < E$ (meV) $< E_F+25$ : (b) SISS (BISS) lead to a sharp (monotonous) energy dependence at $E_F$; and (c-d) spatial charge density maps, taken along the dashed line of panel (a), show how the numerous C and N sites of MnPc exhibit a highly spin-polarised density of states at $E_F$ that, furthermore, are hybridized with Co states and thus contribute to conduction.

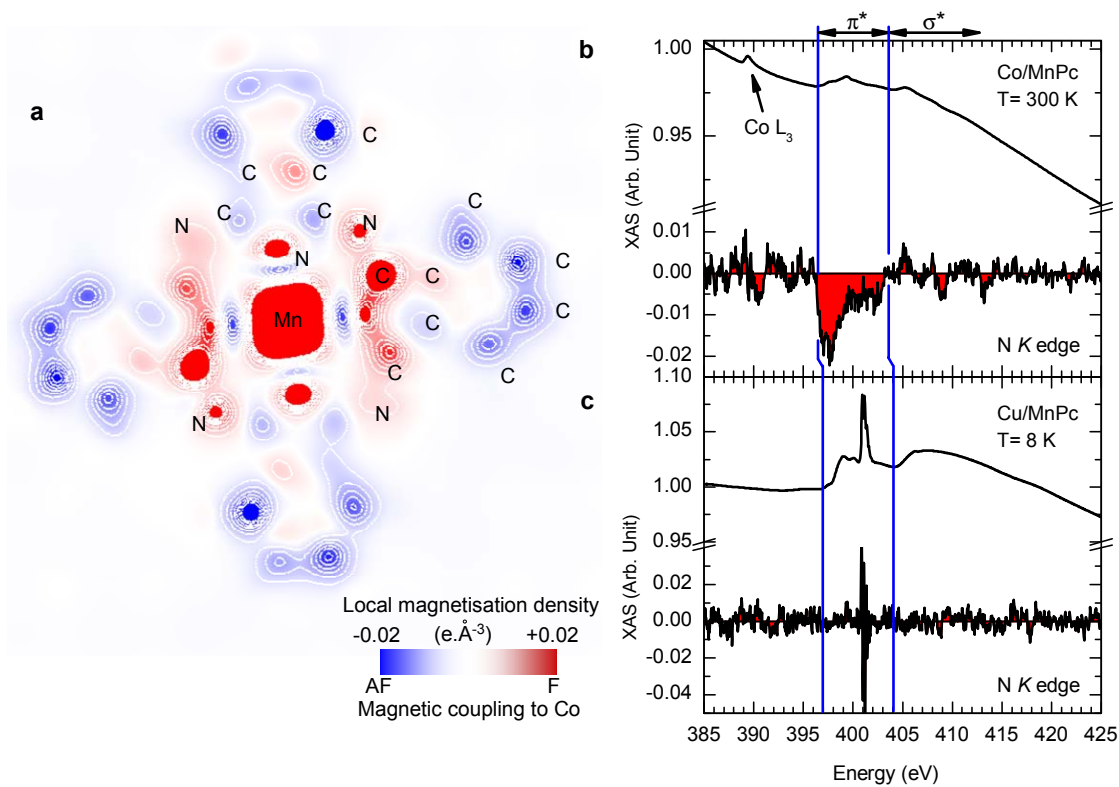

Fig. 4: **Magnetic moments induced through direct exchange onto the molecular sites provide a signature of the Co/MnPc spinterface.** (a) Top view of the on-site magnetisation density of the MnPc molecule adsorbed onto Co. While the pyrrole cage around M is ferromagnetically coupled to Co (F, red), that of the C-based benzene rings is mostly coupled antiferromagnetically (AF, blue) to Co. x-ray magnetic circular dichroic spectra acquired for H=5T and a 45° angle of photon incidence to the sample surface reveal a magnetic polarisation of the N π states of MnPc for (b) Co/MnPc(0.5ML) at T=300K but not (c) Cu/MnPc(1.2ML) even at T=8K. This confirms that the z-DOS of N just above $E_F$ is spin-polarised. The slight energy shift of the N edge onset when going from Cu to Co reflects an increase in chemisorption strength[6].

# Direct room-temperature observation of a highly spin-polarised spinterface

F. Djeghloul[1], F. Ibrahim[1], M. Cantoni[2], M. Bowen[1], L. Joly[1,3], S. Boukari[1], P. Ohresser[4], F. Bertran[4], P. Lefèvre[4], P. Thakur[5], F. Scheurer[1], T. Miyamachi[6], R. Mattana[7], P. Seneor[7], A. Jaafar[1], C. Rinaldi[2], S. Javaid[1], J. Arabski[1], J.-P. Kappler[1], W. Wulfhekel[6], N.B. Brookes[5], R. Bertacco[2], A. Taleb-Ibrahimi[4], M. Alouani[1], E. Beaurepaire[1], W. Weber[1]

**Supplementary Information**

In this Supplementary Information section, we provide additional data in support of our demonstration within the main text of large values of spin polarization at the Fermi level for electron current propagating perpendicularly to the interface between Co and MnPc or $H_2Pc$. We discuss the analysis methodology that was used in order to extract the photoemission signal resulting only from molecular sites.

**Spin-polarized Photoemission Analysis of individual molecular layers: A Methodology**

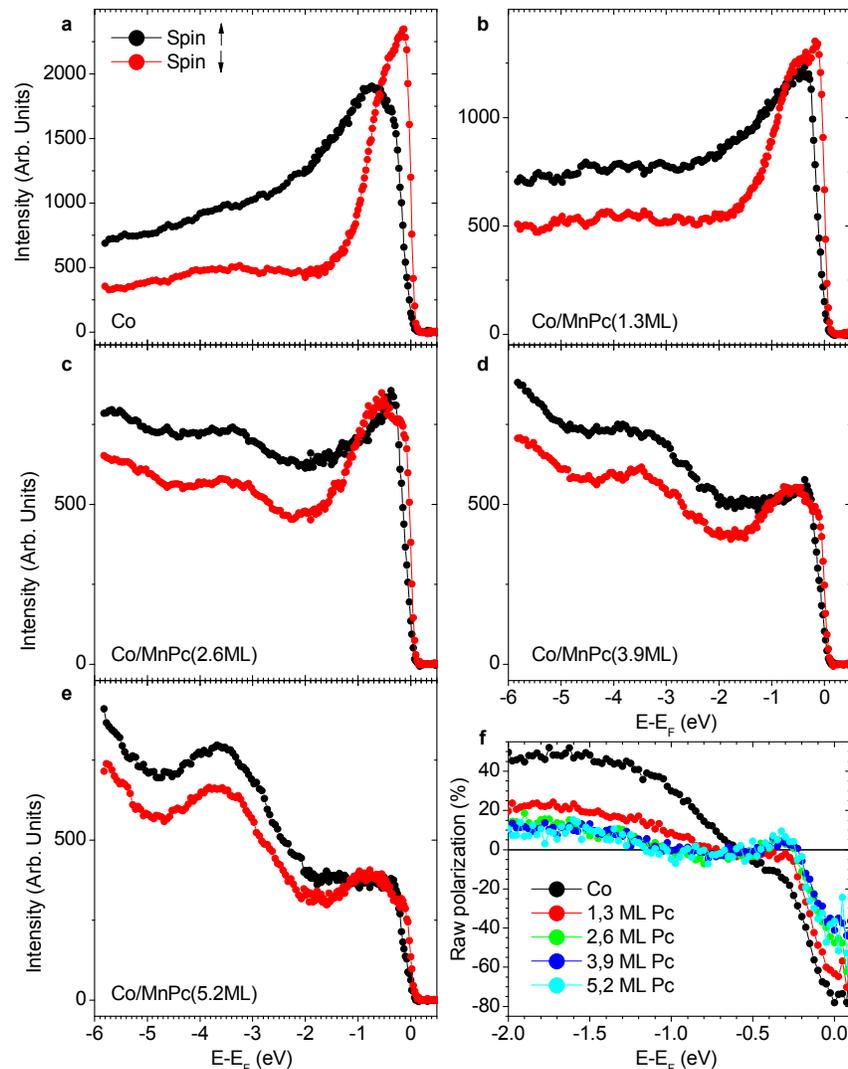

Fig. S1: Spin-polarized photoemission spectra (spin up: black symbols, spin down: red symbols) of Co(001)/MnPc for five different MnPc thicknesses and the spin polarization as a function of binding energy. The photon energy is 20 eV.

We present in Fig. S1 the raw direct photoemission data for both spin channels, as well as the resulting spin polarization P, for the Cu(001)//Co(15ML)/MnPc(xML) system (x = 0, 1.3, 2.6, 3.9, and 5.2). The absolute intensities can be compared as all measurements were performed under the same conditions, in particular with the same incoming photon intensity. As MnPc thickness increases, there appears near $E_F$, against the strong backdrop of Co-induced negative P, a relative decrease in this negative P. Since this MnPc-induced upturn in fact leads to a switchover in sign, we infer that additional MnPc is not merely suppressing the negative P of Co, but is in fact contributing a net positive P. Given the much larger density of states on the Co sites relative to that of the single Mn site, the N and the C sites, this positive P contribution must be substantial and be weighed more heavily within the overall signal since it is at the sample surface.

To extract the spin polarization P of only the molecular sites, we adopt a subtraction procedure that takes into account the attenuation of the signal arising from ever deeper atomic sites away from the sample surface. Since our theory shows that the spin-hybridized interface states occur within only a few eV from the Fermi level, one element of this procedure is to ensure that, once the molecular signal is extracted, the spin polarization far from $E_F$ is zero.

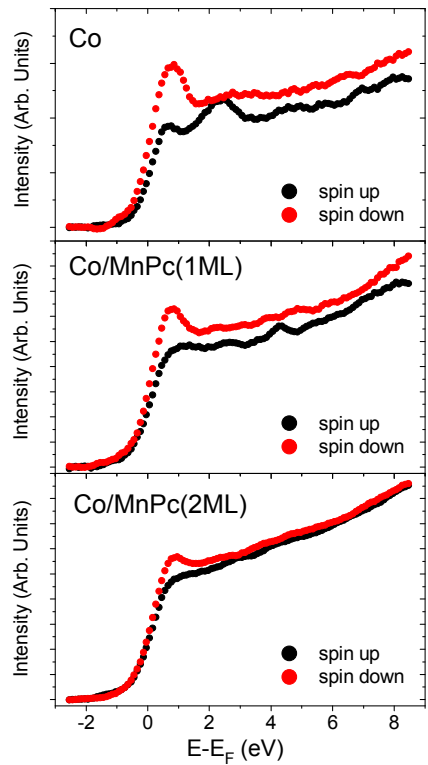

Fig. S2: Spin-polarized inverse photoemission spectra (spin up: black symbols, spin down: red symbols) of Co(001)/MnPc for three different MnPc thicknesses (0, 1, and 2 ML) as a function of binding energy. Photons with an energy of 9.3 eV are detected.

The following subtraction procedure is described regarding spin-polarized direct photoemission data. A similar procedure is applied regarding spin-polarized inverse photoemission data (see Fig. S2). Our methodology supposes that the spin-resolved photoemission signal arising from Co sites remains essentially unchanged when the Co film is covered by MnPc. It also rests upon the observation that molecular adsorption onto Co clearly has a strong impact on photoemission, even in the raw spectra, implying substantial, not just second-order, changes. Several aspects support this supposition. 1) Supposing an inelastic mean free path in Co of about 0.8 nm, only 20 % of the Co photoemission signal is coming from the topmost layer while 80% are coming from the lower Co layers. Thus, the strong changes that we witness are not likely to arise from modifications of the topmost Co layer upon Pc adsorption. 2) Spin-resolved spectra at much higher photon energy (100 eV), which are dominated by Co 3d structures due to the strong photon-energy dependence of the photoionization cross section, show strong similarity for the uncovered and the 2.6 ML MnPc-covered Co film. 3) To check how the magnetic properties of Co are affected by the adsorption of MnPc, we performed x-ray magnetic circular dichroism measurements at the Co L edges at room temperature. Using the so-called sum rules, we find 1.73 µB for the average spin moment per Co atom for 3 Co ML/Cu(001). Upon adsorption of 1 ML MnPc, the Co average spin moment is reduced to 1.67 µB/at. If we assume that this reduction is borne only on the interface sites, then these sites carry a spin moment of at least 1.55 µB /at. This minor reduction in spin moment is confirmed by ab-initio DFT calculations including spin-orbit coupling. This clearly excludes the possibility of a Co magnetic dead layer, and shows that modifications to the Co surface sites are a secondary effect. Such a small loss of magnetization cannot account for the large changes in spin-polarized photoemission that we observe in the raw data. We therefore assert that, from our complementary photoemission, XAS experiments, and with theoretical support, that any reduction in the magnetic moment of Co sites that hybridize with molecular sites cannot account for the large signal that we witness in the subtracted photoemission data.

In order to obtain the spin-resolved molecule-induced photoemission spectra $f_{up,down}(MnPc(xML))$, the pure Co spectra $f_{up,down}(Co)$ have therefore to be subtracted from the Co/MnPc(xML) spectra $f_{up,down}(Co/MnPc(xML))$ by taking into account the attenuation of the Co signal due to the coverage of xML MnPc:

$f_{up,down}(xML\ MnPc) = f_{up,down}(Co/MnPc(xML)) - \exp(-x/\lambda)\ f_{up,down}(Co)$

with $\lambda$ the inelastic mean free path of the electrons in MnPc. Unfortunately, this "straight-forward"-procedure has the great disadvantage that both the MnPc thickness and $\lambda$ have to be known with great accuracy. However, in particular the values of $\lambda$ reported in the literature for low kinetic energies are significantly scattering such that a reliable determination of for instance the interfacial spin polarization becomes difficult. Moreover, the knowledge of the absolute MnPc thickness is necessary. We nevertheless try in a first step to determine the difference spectra by taking the absolute MnPc thicknesses as granted and by supposing a "reasonable" range of $\lambda$ values (between 0.8 and 1.6 nm). Application of this method to both direct and inverse spin-resolved photoemission reveals that the intensity of the difference spectra around $E_F$ is significantly and positively polarized.

In a second step we try to obtain the above attenuation factor without being obliged to know neither the absolute MnPc thickness nor the value of $\lambda$. The following procedure is only based on the fact, that the relative MnPc thicknesses are known and that the experiments are performed under the same conditions (same photon intensity). We emphasize that the relative thicknesses could be determined with great accuracy, while the absolute thicknesses are affected by a larger margin of error.

First, we plot the spin-resolved intensity as a function of MnPc thickness for several binding energies (see Fig. S3). For the spin-up intensity an exponential decrease is seen for energies below -0.4 eV. For lower binding energies down to the Fermi level clear deviations from the exponential behaviour are observed indicating the presence of a spin-up hybridized interface state in this energy range. For higher binding energies of several eV (not shown), on the other hand, the intensity is rather increasing with thickness indicating the appearance of molecular (bulk) states. For the spin-down intensity an exponential decrease is found for energies between 0 and -0.5 eV, while strong deviations from this behaviour are seen for higher binding energies. This indicates the presence of a spin-down hybridized interface state in this higher binding energy range. In fact, theory expects that only the first ML contributes to the DOS by the creation of interface states within this energy range while all other layers contribute only very little. For a given binding energy, we may extract the attenuation factor by taking the ratio of intensity before and after deposition of MnPc. For which binding energy and spin direction should this ratio be calculated? From the above discussion of the spin-polarized intensity as a function of MnPc thickness for several binding energies it is clear that we have to take the spin-down intensity in the energy region close to the Fermi level. In fact, in this energy region the spin-down intensity shows no or at least very weak MnPc-induced intensity, such that only spin-down Co features are seen which are attenuated by the MnPc layers. Thus the ratio of the spin-down intensities close to $E_F$ yields directly the attenuation factor necessary for the subtraction procedure. To ensure that the attenuation factor is valid, we slightly varied it to see how it influences the spin polarization of **the difference** spectra (**!**) at higher binding energies (from -4.5 to -6 eV). For the difference spectra $f_{up,down}$(1.3ML MnPc), for example, zero polarization is found for the same factor which we find also from the exponential intensity decrease (see Fig. S4).

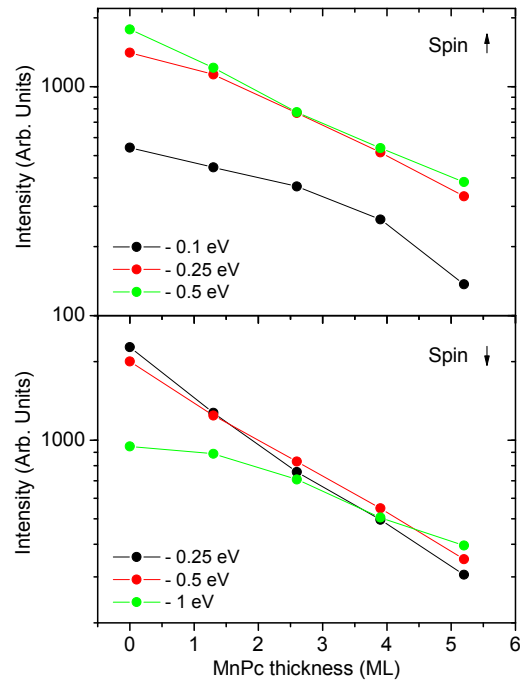

Fig. S3: The photoemission intensity in the spin ↑ (top) and the spin ↓ channel (bottom) for different binding energy positions as a function of MnPc thickness. The photon energy is 20 eV. The attenuation factor used to calculate the difference spectra $f_{up,down}$(1.3ML MnPc), for instance, is obtained by taking the ratio of the intensities for 1.3 ML MnPc and pure Co in the spin-down channel at low binding energies (-0.25 eV): α = 1274/2270 = 0.56. Note the logarithmic intensity scale.

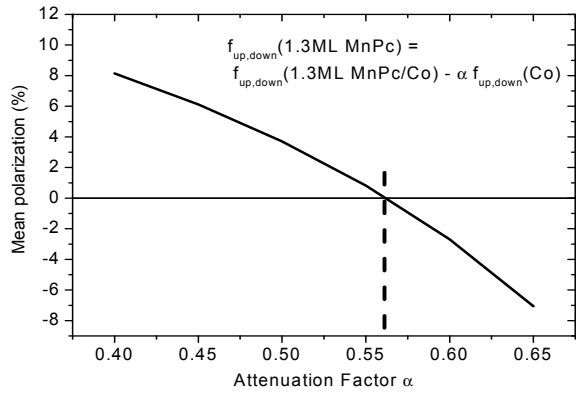

Fig. S4: The mean spin polarization in the binding energy range from -4.5 to -6 eV of the difference spectra $f_{up,down}$(1.3ML MnPc) as a function of the attenuation factor $\alpha$. Zero polarization is obtained for $\alpha = 0.56$.